\documentclass[preprint,showpacs,preprintnumbers,amsmath,amssymb]{revtex4}

\usepackage{graphicx}
\usepackage{dcolumn}
\usepackage{bm}

\begin{document}


\title{Slow dynamics of
salol: a pressure and temperature dependent \\
light scattering study.}
\author{
    L.~Comez$^{1,2}$, S. Corezzi$^{1}$, D. Fioretto$^{1}$, H. Kriegs$^{2}$, A. Best$^{2}$, and W. Steffen$^{2}$}
\address{
     $^1$
     Dipartimento di Fisica and INFM, Universit\`a di Perugia,
     I-06123, Perugia, Italy. \\
     $^{2}$
     Max Planck Institute for Polymer Research, Ackermannweg 10,
     D-55128, Mainz, Germany.
     }
\date{\today}

\begin{abstract}
We study the slow dynamics of salol by varying both temperature
and pressure using photon correlation spectroscopy and
pressure-volume-temperature measurements, and compare the
behavior of the structural relaxation time with equations derived
within the Adam-Gibbs entropy theory and the Cohen-Grest free
volume theory. We find that pressure dependent data are crucial
to assess the validity of these model equations. Our analysis
supports the entropy-based equation, and estimates the
configurational entropy of salol at ambient pressure $\sim$ 70\%
of the excess entropy. Finally, we investigate the evolution of
the shape of the structural relaxation process, and find that a
time-temperature-pressure superposition principle holds over the
range investigated.

\end{abstract}


\maketitle
\section{INTRODUCTION}
\label{introduction}

The study of the supercooled liquid and glassy states in molecular
systems is, nowadays, one of the most important topics in the
physics of disordered materials. Though the molecular processes
underlying glass formation still constitute an unsettled subject,
some traces of universality in the behavior of highly viscous
liquids near vitrification have been noticed. As general
characteristics, on approaching the glass transition the
structural ($\alpha$) relaxation process shows {\it i}) a
non-Debye behavior of the relaxation function, and {\it ii}) a
dramatic increase of the relaxation time, $\tau$.

Different physical routes can be covered to get vitrification.
Decreasing the temperature, $T$, is the common way to form a
glass. However, varying the pressure, $P$, also represents an
effective means. Indeed, the effects on molecular motions of an
isothermal compression resemble those which are caused by an
isobaric cooling. For practical reasons, cooling is generally
preferred, since high pressures (of the order of MPa) are
necessary to produce dynamical changes similar to those obtained
by changing the temperature within few tens of degrees. Anyway,
the study of the $\alpha$ relaxation pressure dependence can give
an insight into the nature of the liquid-glass transition.

The past few years have actually seen a growing use of
hydrostatic pressure in experimental investigations of glass
formers (see for instance,
\cite{Corezzi99,Paluch98,PaluchPRL2000,Casalini,CasaliniJCP02,
PaluchJCP03a,PaluchJCP03b,PaluchJCP02a,PaluchJCP02b}). Such
experiments provide a more stringent testing-ground for the
numerous models proposed of the structural relaxation time
evolution near vitrification. Among these, two are the most
widely used, which are based on the free volume and
configurational entropy concepts. Free volume approaches consider
the decrease of unoccupied volume as the basic mechanism leading
to structural arrest of a liquid system. The alternative view is
that the progressive slowdown of molecular motions responsible
for the glass transition is due to a reduction of the system's
configurational entropy.

In this paper, we test on salol the ability of free volume and
configurational entropy models to interpret the temperature and
pressure dependence of the structural relaxation time. Salol, is a
good candidate since much of the thermodynamic data is known,
allowing refinement on testing theoretical models. It has
intensively been studied at ambient pressure with several
spectroscopic techniques, like Brillouin scattering
\cite{Dreyfus92}, depolarized light scattering
\cite{Li92,Cummins93,Krakoviack97}, impulsive stimulated light
scattering \cite{NelsonPRL95}, optical Kerr effect spectroscopy
\cite{Hinze00}, neutron scattering \cite{Toulouse93}, x-ray
diffraction \cite{Eckstein00}, and dielectric spectroscopy
\cite{Stickel95}. On the other hand, few experiments have been
carried out by varying both temperature and pressure, namely
dielectric spectroscopy \cite{CasaliniJPC03}, depolarized Raman
scattering \cite{Pratesi00}, and viscosity measurements
\cite{Schug80}. Recently, some of us presented a preliminary
investigation \cite{ComezPRE02} on salol performed in the $T$ and
$P$ domain by using photon correlation spectroscopy. Here, we
extend our analysis through pressure-volume-temperature (PVT)
data taken in both the supercooled and crystalline phases. We
show how an appropriate use of the PVT results provides a
realistic estimate of the configurational contribution to the
excess entropy of salol. Finally, we compare our $\tau(T,P)$ data
with the prediction of the pressure-extended Cohen-Grest model
\cite{CohGre79}, derived in the frame of the free volume theory.

\section{THEORY}
\label{THEORY}

\subsection{THE PRESSURE EXTENDED ADAM-GIBBS (PEAG) MODEL}
\label{PEAG model}

The entropy model by Adam and Gibbs (AG) \cite{AG} is based on the
concept of configurational entropy and the assumption of
cooperatively rearranging regions. Starting from the observation
that the sluggish relaxation behavior governing the glass
transition is a manifestation of a dearth of configurations
accessible to the system, the AG theory states a relationship
between the structural relaxation time, $\tau$, and the
configurational entropy $S_c$:

\begin{equation}
  \tau=\tau_0 \exp \left( \frac{C_{AG}}{T S_c} \right),
  \label{AG}
\end{equation}
where $C_{AG}$ is nearly constant, and $\tau_0$ is the relaxation
time at very high temperature. $S_c$ measures the entropic
contribution arising from the possibility of a system to
rearrange its structure in different configurations, which is
typical of a liquid. Theoretically, a quantitative evaluation of
$S_c$ can be done in terms of the difference between the entropy
of the liquid phase and the entropy of an ideal amorphous-solid
phase (ideal glass) in which only vibrations (harmonic and
anharmonic) and secondary relaxation processes are active
\cite{GoldsteinJCP76,JohariJCP2000}. This quantity can, in
principle, be determined by computer simulations, but is
inaccessible to experiments in a direct manner. Some efforts have
been made to bypass a direct experimental determination of
configurational entropy in a number of liquids. Unfortunately,
the procedures proposed require an independent estimate of
vibrational contributions to the entropy over a broad range of
temperatures \cite{Phillips89,giapponesi} or an estimate of the
excess vibrational entropy at $T_{g}$ \cite{Johari}, all of which
implying non-trivial approximations. We also remark that all the
previous estimates of $S_c$ are based on temperature dependent
data alone, and are not constrained by pressure dependent data.\\
Furthermore, much literature documented the extensive use of the
experimentally accessible liquid over crystal (or glass) excess
entropy, $S_{exc}$, in place of $S_{c}$, showing that the AG
expression works well in a number of systems with $S_{c}$
replaced by $S_{exc}$ \cite{GreetTurnbull67,RicAng98}. In this
context, understanding the relationship between $S_{exc}$ and
$S_{c}$ is a challenging issue. A proportionality of these two
quantities at atmospheric pressure has recently been proposed
\cite{Martinez}, but a verification of such hypothesis through a
relaxation experiment performed as a function of temperature
alone cannot be conclusive, as the proportionality constant would
simply renormalize the value of $C_{AG}$ in Eq. (\ref{AG}).

Building on this background, a method based on a pressure extended
Adam-Gibbs (PEAG) equation has recently been proposed by some of
us \cite{Corezzicond-mat} to analyze temperature and pressure
dependent relaxation measurements. The pressure dependence of
$S_{c}$ has been introduced in Eq. (\ref{AG}) writing the
configurational entropy of a system at a given $T$ and $P$ as a
sum of (i) an isobaric contribution at zero pressure,
$S_{c}^{isob}(T,0)$, and (ii) an isothermal contribution at
temperature $T$, $S_c^{isoth}(T,P)$:

\begin{equation}
 S_c(T,P)=S_{c}^{isob}(T,0)+S_c^{isoth}(T,P)\\ \label{TECE}
\end{equation}

\noindent (i) Here, the isobaric configurational term, at zero
pressure, is assumed proportional to the excess entropy:
\begin{equation}
  S_c^{isob}(T,0)=\Phi S_{exc}^{isob}(T,0).
  \label{Scphi}
\end{equation}
The parameter $\Phi$ ($\leq$1) quantifies the fraction of excess
entropy at $P$=0 arising from structural configurations. In
addition, the excess entropy contains any contribution from
secondary relaxation processes and vibrational motions
\cite{GoldsteinJCP76,JohariJCP2000,Fischer02}. It can be evaluated
from the heat capacity of the liquid and the crystal, through the
equation:
\begin{eqnarray}
S_{exc}^{isob}(T,0)&=&S^{liquid}(T)\!-S^{crystal}(T)\nonumber\\
&=&\Delta
S_{f}-\int_{T}^{T_{f}}\!\!\!\left(C_{p}^{liquid}\!\!-\!C_{p}^{crystal}\right)/T'dT'
\label{SmeltScrystal}
\end{eqnarray}
where $\Delta S_{f}\!=\!\Delta H_{f}/T_{f}$ is the entropy of
fusion.

\noindent (ii) According to the Maxwell relationship
$\left(\partial S/\partial P \right)_{T}=-\left(\partial
V/\partial T\right)_{P}$, the isothermal term in Eq.(\ref{TECE})
can be written

\begin{eqnarray}
S_c^{isoth}(T,P)= -\int_0^P [\Delta\left(\partial V/\partial
T\right)_{P}] dP' \label{Scisoth}
\end{eqnarray}
where $\Delta\left(\partial V/\partial T\right)_{P}=\left(\partial
V/\partial T\right)_{P}^{liquid}-\left(\partial V/\partial
T\right)_{P}^{non-struct}$ is the configurational thermal
expansion at temperature $T$ \cite{expansion}. This term can be
evaluated from PVT measurements as follows. The Tait equation
\cite{Tait} is used to describe the volume of the liquid phase as
a function of $T$ and $P$
\begin{equation}
V^{liquid}(T,P)=V^{liquid}(T,0)\left[1-C  \ln\left(
1+P/B\right)\right], \label{Tait}
\end{equation}
where $C$ is a dimensionless constant, and $B(T)$ is well
described by $B(T)=b_1 \exp(-b_2T)$, where $b_1$ has the
dimension of pressure and $b_2$ of inverse of temperature
\cite{VanKrevelen}. Moreover, it is reasonable to presume that
the pressure dependence of the thermal expansion of the ideal
glass would be much smaller than that of the liquid, and can be
neglected. Accordingly, the non-structural thermal expansion at
$P$ can be replaced by its value at $P$=0, i.e., $\left(\partial
V/\partial T\right)_{P}^{non-struct}\approx \left(\partial
V/\partial T\right)_{0}^{non-struct}$. Under these assumptions,
calculating the integral in Eq. (\ref{Scisoth}) yields

\begin{widetext}
\begin{equation}
S_c^{isoth}(T,P)\approx-\left( \frac{\partial V}{\partial T}
\right)_0^{liquid} \left[ P+ hCP - BC  \left(h +\frac{P}{B}
\right) ln \left( 1+\frac{P}{B} \right) \right] + P \left(
\frac{\partial V}{\partial T} \right)_0^{non-struct}
\label{wideeq}
\end{equation}
\end{widetext}
where $h=1-b_2/\alpha$, and $\alpha=1/V( \partial V /\partial T
)_0$ is the thermal expansion coefficient at zero pressure.

In conclusion, combining Eqs.~[\ref{AG}-\ref{Scphi}] provides a
formula for the structural relaxation time as a function of
temperature and pressure:
\begin{equation}
  \tau(T,P)=\tau_0 \exp \left[ \frac{C_{AG}}{T(\Phi S_{exc}^{isob}+S_c^{isoth})}
  \right],
  \label{AGphi}
\end{equation}
with $ S_{exc}^{isob}$ and $S_c^{isoth}$ given by Eq.
(\ref{SmeltScrystal}) and (\ref{wideeq}), respectively. It is
important to emphasize that the expression of $S_c^{isoth}$, Eq.
(\ref{wideeq}), prevents the parameter $\Phi$ in
Eq.~(\ref{AGphi}) from playing the role of a simple
renormalization constant.

\subsection{THE PRESSURE EXTENDED COHEN-GREST (CG) MODEL}
\label{CG model}

Within a free volume picture, Cohen and Grest \cite{CohGre79}
derived a model to describe the behavior of dense liquids and
glasses on the basis of a percolative approach. The existence is
assumed of glass-like and liquid-like domains. The fraction, $p$,
of these latter increases with temperature, and a percolative
(infinite) cluster does exist above a critical concentration
$p_c$, at which the transition from the glass to the liquid state
occurs. The model predicts an analytical expression for the free
volume $v_f$ which is valid in a broad range of temperatures:
\begin{equation}
  v_f=\frac{k}{2\xi_0}\lbrace{T-T_0+[(T-T_0)^2+4v_a\xi_0T/k]^{1/2}\rbrace}
  \label{vf}
\end{equation}
where $T_0$, $\xi_0$, and $v_a$ are constants with the dimension
of temperature, pressure, and volume, respectively. For $p$ near
and greater than $p_c$, a link is established between $v_f$ and
the diffusion coefficient $D$, which recovers the Doolittle
equation \cite{doolittle}, $D=D_0p \exp(-v_m/v_f)$, in the case of
$v_m/v_f<<\bar{\nu}$. Here, $v_m$ is the molecular volume, $D_0$
is a constant, and $\bar{\nu}$ is the average size of the
liquid-like clusters. A similar result is assumed for the
rotational correlation time, $\tau=\tau_0 \exp(v_m/v_f)$
\cite{CohGre81}, where $\tau_0$ is the value of $\tau$ in the
limit of very high temperature under isobaric conditions. On this
basis, a simple equation for the structural relaxation time in
the supercooled state can be written:
\begin{equation}
 \log\tau(T)=A_{CG}+\frac{B_{CG}}{T-T_0+[(T-T_0)^2+C_{CG}T]^{1/2}}\label{tauCG}
\end{equation}
where $A_{CG}$ is related to the pre-exponential factor $\tau_0$,
and the parameters $B_{CG}=2 \xi_0 v_m \log e /k$ and
$C_{CG}=4v_a\xi_0T/k$ have the dimension of temperature, and must
assume positive values to have a physical meaning.

Cohen and Grest incorporate the effect of pressure in their
theory by including an additional term, proportional to pressure,
into their expression for the local free energy. As a
consequence, the pressure dependence of the relaxation time can
be obtained by changing $\xi_0 \longrightarrow \xi_0 +P$. The
temperature parameter $T_0$ is also affected by this change, via
the relationship $kT_0=kT_1+v_a\xi_0$, with $T_1$ a constant,
which yields $T_0(P)=T_0+(v_a/k)P$. The final expression for
$\tau(T,P)$ is:
\begin{equation}
 \log\tau(T,P)=A_{CG}+\frac{B_{CG} D_{CG}}{T-T_0^*+[(T-T^*_0)^2+C_{CG} D_{CG} T]^{1/2}}\label{tauCGex}
\end{equation}
with $D_{CG}=1+P/\xi_0$ and $T_0^*=T_0-(C/4\xi_0)P$. Note that
this expression contains five parameters, i.e. $A_{CG}$, $B_{CG}$,
$T_0$, $C_{CG}$, and $\xi_0$, only the first four appearing in the
temperature dependent expression at $P$=0, i.e. in Eq.
(\ref{tauCG}).

\section{Experiment}
\label{experiment}

\subsection{PVT Measurements}
\label{PVT}

\begin{figure}[t]
\includegraphics[width=9.0cm]{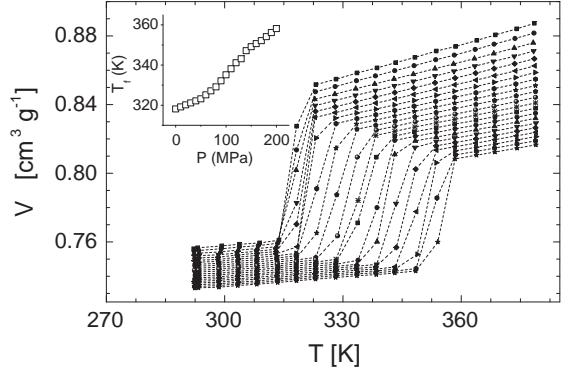}
\caption{\label{volume-SALOL} Temperature dependence of the
volume of salol in the crystal and liquid state at different
pressures. The pressures are, from top to bottom, from 0.1 MPa to
200 MPa in steps of 10 MPa. In the inset the melting temperature
versus pressure deduced from the PVT measurements here reported.}
\end{figure}

Measurements of specific volume change $\Delta V(T,P)$ of
crystalline and liquid salol were taken in an isothermal mode of
operation by using a confining fluid technique \cite{ZolWal95}.
The PVT data were acquired on a GNOMIX apparatus \cite{Gnomix}
described in Refs. \onlinecite{ZolWal95,ZolBol76}. The sample
(salol) and the confining fluid (mercury) were contained in a
rigid sample cell. A thin nickel foil sample cup surrounding the
sample was used to guarantee hydrostatic pressure during the
experiment. Silicon oil was used as pressurizing fluid. The
temperature was recorded (for operational reasons) close to the
sample, but actually in the pressurizing silicon oil. At a fixed
temperature, starting from the low-temperature end, pressure was
increased to 200 MPa, and data were recorded in pressure
intervals of 10 MPa. On completion of measurements along one
isotherm, the temperature setting was increased 5 K higher, and
the pressure measurements were repeated. $\Delta V(T,P)$
measurements were converted into specific volume $V(T,P)$ data by
using a reference density value, $\rho$=1.1742 g cm$^{-3}$ at
$T$=323.15 K. The whole set of PVT measurements between $T$=290 K
and 380 K over the 0.1-200 MPa range of pressure is reported in
Fig. \ref{volume-SALOL}. The step in the data at a given pressure
marks the fusion/crystallization temperature.

\subsection{Photon correlation measurements}
\label{PCS}

Photon correlation spectroscopy (PCS) measurements under high
hydrostatic pressure, up to 190 MPa, were taken at different
temperatures (namely 267.1, 268.6, 271.0, 274.6, 278.3 and 280.4
K). Depolarized (VH) light scattering spectra were collected in
the 90$^{\circ}$ geometry using an apparatus consisting of an
Ar-ion laser, operating at 514.5 nm, a home made thermostated high
pressure cell (a detailed description of the cell is reported in
refs. \onlinecite{FytPat82,Fytas84}), and an ALV5000E digital
correlator. The scattered light was collected by a single mode
fiber optics and detected by an avalanche diode (Sandercock).
High pressure was generated by using nitrogen pressurized by a
Nova Swiss membrane compressor and introducing the gas over steel
capillaries connected with the high pressure cell. The pressure
was measured by a Heise gauge with a resolution of 0.3 MPa, and
the temperature by a thermocouple with a typical error of 0.1 K.
Special care was taken to prepare the sample avoiding
crystallization on both lowering the temperature and increasing
the pressure. A cleaning procedure to have dust-free cells was
used consisting of rinsing the cells with freshly distilled hot
acetone. Salol [2-hydroxy benzoic acid phenyl ester, 2-(HO)
C$_{6}$H$_{4}$CO$_{2}$C$_{6}$H$_{5}$] purchased from Aldrich
company, purity 99 \%, was filtered (0.22$\mu$m Millipore filter)
into the dust-free cylindrical cell of 10 mm o.d. at about
80$^{\circ}$C. The sample was then brought back to room
temperature at a very slow cooling rate. The measurements were
performed following isothermal curves by varying the pressure.
Each isothermal run was usually done from the higher to the lower
value of pressure, this procedure assuring a shorter
equilibration time before starting the measurement. Finally, we
checked that the diffusion time of N$_2$ was long enough to
prevent contamination of the scattering volume during the
experiment. To this end the forward beam was continuously
monitored on a black screen to directly visualize possible
vertical gradients of the refractive index of the sample.

\begin{figure}[t]
\includegraphics[width=8.2cm]{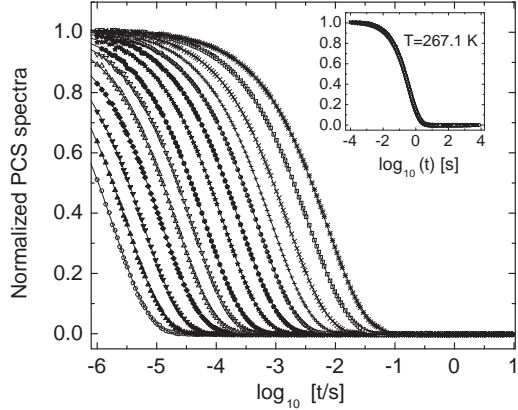}
\caption{\label{Fig_spettri} Normalized photon correlation
functions collected at a constant temperature of 267.1 K.
Pressures from left to right are 88, 95, 102.5, 110.5, 119, 125,
132.5, 141, 148.5, 156.5, 163.5, 171, 181, and 189.5 MPa. The
solid lines represent the fits to the data using the KWW
function. The isothermal spectra at 267.1 K taken at different
pressures rescale on a master curve as shown in the inset.}
\end{figure}

\section{RESULTS}
\label{results}

\subsection{Thermodynamic parameters}
\label{PVT results}

The $T$ and $P$ dependence of the volume can be expressed through
the Tait equation, Eq. (\ref{Tait}), found to be valid for a wide
range of materials including liquids and polymers, for changes of
the volume up to 40 $\%$ of the initial value. From the analysis
of the data at atmospheric pressure in the liquid state we
numerically find a constant value of the thermal expansion
coefficient $\alpha=\left(\partial V/\partial
T\right)_{0}^{liquid}\!\!/V^{liquid}(T,0)$, consistent with the
expression $V^{liquid}(T,0)=V_{0}\exp(\alpha T)$ describing the
temperature behavior of the volume of liquid salol at $P=0$
\cite{zeropressure}. The whole set of PVT data in the liquid
state is then fitted by Eq.~(\ref{Tait}). In Fig. 3 the
experimental data are shown together with the result of the fit
(solid lines). An excellent agreement between experimental points
and fit curves is obtained with the values of the parameters
$V_0$, $\alpha$, $b_1$, $b_2$, and $C$ reported in Tab.
\ref{tablePVT}. It is possible to recognize some generality of
the parameters of the Tait equation \cite{VanKrevelen}. Indeed,
the values of $C$ ($\sim 0.09$) and $b_2$ ($\sim$ 4x10$^{-3}$
K$^{-1}$) have been found to be almost the same for a lot of
materials, liquids and polymers, including chlorinated biphenyl
(PCB62) \cite{CasaliniJCP02}, diglycidylether of bisphenol A
(DGEBA) \cite{PaluchJCP03a}, bis-phenol-C-dimethylether (BCDE) and
bis-kresol-C-dimethylether (BKDE) \cite{PaluchJCP03b},
phenylphthalein-dimethylether (PDE) \cite{PaluchJCP02b} and
cresolphthalein-dimethylether (KDE) \cite{PaluchJCP02a}.\\
In the crystalline phase, PVT measurements allow us to evaluate
the thermal expansivity at different pressures. In particular, we
find that $\left(\partial V/\partial T \right)_{P}^{crystal}$
ranges from about $4.5\times 10^{-8}$ \mbox{m$^{3}$ mol$^{-1}$
K$^{-1}$} at $P$=0.1 MPa to about $3.5\times 10^{-8}$
\mbox{m$^{3}$ mol$^{-1}$ K$^{-1}$} at $P$=200 MPa, with an
average value $\left(\partial V/\partial T
\right)_{\bar{P}}^{crystal}\sim (4.0\pm 0.5)\times 10^{-8}$
\mbox{m$^{3}$ mol$^{-1}$ K$^{-1}$} over the pressure range
investigated.

\begin{figure}[t]
\includegraphics[width=9.0cm]{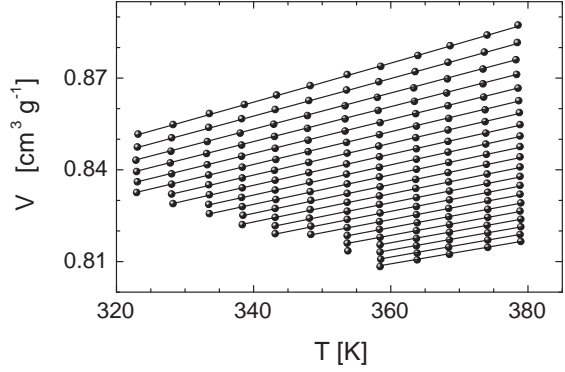}
\caption{\label{fig:volume-SALOL} Temperature dependence of the
volume of salol in the liquid state. The solid lines through
symbols are the best fit with the Tait equation of state,
Eq.~(\ref{Tait}), with $V^{liquid}(T,0)=V_{0}\exp(\alpha T)$ and
$B(T)=b_{1}\exp(-b_{2}T)$.}
\end{figure}

\begin{figure}[t]
\includegraphics[width=8.5cm]{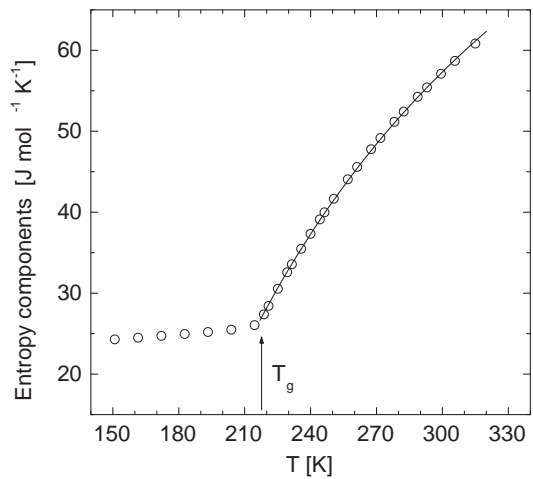}
\caption{\label{figSexcSALOL} Temperature dependence of the
excess entropy over crystal,
\mbox{$S_{exc}\!=\!S^{melt}\!\!-S^{crystal}$}, calculated from the
calorimetric data.
 The solid
line represents the fit of the experimental data according to
\mbox{$S_{\infty}\!\!-k/T$}.}
\end{figure}

The heat capacity $C_{p}$ of crystalline, glassy, supercooled and
stable liquid salol at atmospheric pressure was measured by
adiabatic calorimetry \cite{Oguni,OguniPrivate}. From these data,
the glass transition temperature $T_{g}$=217$\pm$1 K and the
temperature of fusion $T_{f}$=315.0 K are determined, and the
excess entropy of the liquid over the crystal, $S_{exc}(T)$, is
calculated by using Eq.~(\ref{SmeltScrystal}), with the value
$\Delta S_{f}\!=\!\Delta H_{f}/T_{f}=60.83\pm 0.04$ \mbox{J
mol$^{-1}$ K$^{-1}$} for the entropy of fusion. In
Fig.~\ref{figSexcSALOL} the experimental excess entropy is shown
with circles.

\begin{table}
\caption{\label{tablePVT} Thermodynamic parameters from the
analysis of volumetric measurements.}
\begin{ruledtabular}
\begin{tabular}{lc}
$V_{0}$ [m$^{3}$ mol$^{-1}$]& $(143.8\pm 0.1)\times 10^{-6}$ \\

$\alpha$ [K$^{-1}$]& $(7.36\pm 0.02)\times 10^{-4}$  \\

$b_{1}$ [MPa]& $790\pm20$  \\

$b_{2}$ [K$^{-1}$]& $(4.70\pm0.06)\times 10^{-3}$ \\

$C$& $(8.68\pm0.05)\times 10^{-2}$ \\
\end{tabular}
\end{ruledtabular}
\end{table}

\subsection{Dynamic parameters}
\label{PCS results}

In the PCS experiment the homodyne technique is used, which
measures the normalized time correlation function of the
scattering intensity $g^{(2)}(t)=<I(t)I(0)>/<I^2>$. For a
Gaussian process, the intensity autocorrelation function
$g^{(2)}(t)$ is related to the autocorrelation function of the
scattered field, $g^{(1)}(t)=<E(t)E(0)>/<|E(0)^2|>$, through the
Siegert equation \cite{Berne}:

\begin{equation}
  g^{(2)}(t) = (1+f |g^{(1)}(t)|^2)
  \label{G2}
\end{equation}
where $f$ is a constant. The relaxation function of a
glass-forming system is generally broader than a single
exponential, and experimental data are typically represented by
the phenomenological Kohlrausch-Williams-Watts (KWW) function
\cite{KWW}:

\begin{equation}
  g^{(1)}(t) = [g_0 \exp (-(t/\tau_K)^{\beta_K})]
  \label{KWW}
\end{equation}
Therefore, PCS spectra are fitted by using Eqs. (\ref{G2}) and
(\ref{KWW}). The results show an excellent agreement between
experimental data and fit curves. Typical normalized homodyne
correlation spectra $|g^{(1)}(t)|^2$ (symbols), taken at 267.1 K
in the 88-189.5 MPa pressure range, are represented in Fig. 2
together with their KWW fits (solid lines). The values of the
relaxation time $\tau_K$ and of the stretching parameter
$\beta_K$ have been used to calculate the average relaxation time
$\langle\tau\rangle$, through the formula

\begin{equation} \langle\tau\rangle =
\frac{\tau_K}{\beta_K}\Gamma\left(\frac{1}{\beta_K}\right)
\label{tauav}
\end{equation}
where $\Gamma$ is the Euler $\Gamma$-function. The values of
$\langle\tau\rangle$ as a function of pressure at different
temperatures are shown as symbols in Fig. 5.

Following the evolution with $T$ and $P$ of the shape of the
relaxation function, we find that no appreciable variation is
observable on the stretching parameter by changing $T$ and $P$.
This evidence is further supported when a master plot is drawn,
showing that the spectra taken at different pressures collapse
into a single curve (see inset of Fig. 2). Our determination of
the stretching parameter ($\beta_K=0.68\pm0.02$) agrees with
previous results at ambient pressure and low temperatures from PCS
measurements: $\beta_K=0.66\pm0.03$ \cite{Berger96}, and
$\beta_K=0.60\pm0.08$ \cite{SidebottomSorensen}. Different
techniques, such as dielectric spectroscopy
\cite{Stickel95,Berger96} and impulsive stimulated light
scattering \cite{Yang1995}, also found a time-temperature
superposition (TTS) principle to hold in salol at low
temperatures. Remarkably, our results indicate the validity of a
generalized time-temperature-pressure superposition (TTPS)
principle in the slow dynamic regime, and support recent finding
of only a modest broadening of the dielectric $\alpha$ peak with
increasing pressure up to 0.7 GPa \cite{CasaliniJPC03}.\\
Moreover, Olsen et al. \cite{Olse01} recently reinvestigated TTS
at low temperatures for a large number of systems concluding that
a high-frequency slope of the $\alpha$ peak close to -1/2 is
expected whenever TTS applies. To confront with this expectation,
we first evaluate, through the relationship \cite{Lind80}
\begin{equation}\label{betaKeCD}
\beta_K=0.970 \beta_{CD}+0.144   \qquad 0.2 \leq \beta_{CD}\leq
0.6,
\end{equation}
the value of a Cole-Davidson shape parameter, $\beta_{CD}$,
corresponding to our value of $\beta_K$ in the time domain. We
find $\beta_{CD}$=($0.55\pm0.02)$, and then the $\alpha$ peak
actually decays approximately as $\omega^{-1/2}$ at high
frequencies, at any temperature and any pressure considered here.

\section{DISCUSSION}
\label{discussion}

\subsection{Check of the PEAG model}
\label{CheckPEAG}

Our relaxation data are well in the range in which strong
intermolecular cooperativity is expected for salol
\cite{Stickel95,RicAng98,CasaliniJCP2003}. To check the
consistency of the PEAG model with our relaxation data, following
Sec. \ref{PEAG model} we need to determine both the isobaric
contribution at zero pressure and the isothermal contribution at
temperature $T$ of the configurational entropy, Eq.~(\ref{TECE}).
The former contribution is related to the excess entropy of the
liquid over its crystalline phase at ambient pressure,
Eq.~(\ref{Scphi}). The latter is given by Eq.~(\ref{wideeq}).

The temperature behavior of the excess entropy is well described,
over the whole range between $T_g$ and $T_f$, by the function
$S_{exc}\!=\!S_{\infty}\!\!-k/T$, as observed in a number of other
glass formers \cite{RicAng98}. The best fit curve corresponds to
the parameters $S_{\infty}=137.5\pm 0.3$ \mbox{J mol$^{-1}$
K$^{-1}$}, $k=(24.05\pm 0.08)\times 10^{3}$ \mbox{J mol$^{-1}$},
(see Fig. \ref{figSexcSALOL}). Hence, Eq.~(\ref{Scphi}) becomes
$S_{c}^{isob}(T,0)\!=\! \Phi \! (S_{\infty}\!\!-k/T)$, where
$S_{\infty}$ and $k$ are known, and $\Phi$ will be free in the
global fit with Eq. (\ref{AGphi}).

For what concerns the isothermal term, Eq. (\ref{wideeq}), the
expressions $\left(\partial V/\partial T
\right)_{0}^{liquid}\!\!=\alpha V^{liquid}(T,0)$,
$h=1-b_{2}/\alpha$, and $B=b_{1}\exp(-b_{2}T)$ are known from the
analysis of PVT data. Numerical details are reported in
Tab.~\ref{tabPVTb}. The only parameter which couldn't be
determined experimentally is the thermal expansivity
$\left(\partial V/\partial T \right)_{0}^{non-struct}$ associated
with non-structural contributions. Although the value of this
parameter will be derived from the fit, we expect that such a
value should compare well with that calculated in the crystal of
salol, as our sample is grown in a polycrystalline form that
should mimic better than a perfect crystal the vibrational
properties of an ideal amorphous solid.
\begin{table}
\caption{\label{tabPVTb} Thermodynamic parameters in Eq.
(\ref{wideeq}) calculated from PVT measurements.}
\begin{ruledtabular}
\begin{tabular}{ccccc}
$T$& $P$&$|h|$&$(\partial V/\partial T)^{liquid}_{0}$ &$B$ \\
(K) & (MPa) & &(m$^3$mol$^{-1}$K$^{-1}$) & (MPa) \\
\hline
267.1 &  88.0-189.5 & 3.588 & 1.287x10$^{-7}$ &225.1\\
268.6 & 110.0-180.0 & 3.588 & 1.289x10$^{-7}$ &223.5\\
271.0 & 115.5-185.0 & 3.588 & 1.291x10$^{-7}$ &220.9\\
274.6 & 140.0-185.0 & 3.588 & 1.294x10$^{-7}$ &217.3\\
278.3 & 155.5-190.0 & 3.588 & 1.298x10$^{-7}$ &213.6\\
280.4 & 150.0-194.0 & 3.588 & 1.30x10$^{-7}$ &211.5\\
\end{tabular}
\end{ruledtabular}
\end{table}

\begin{figure}[t]
\includegraphics[width=8.2cm]{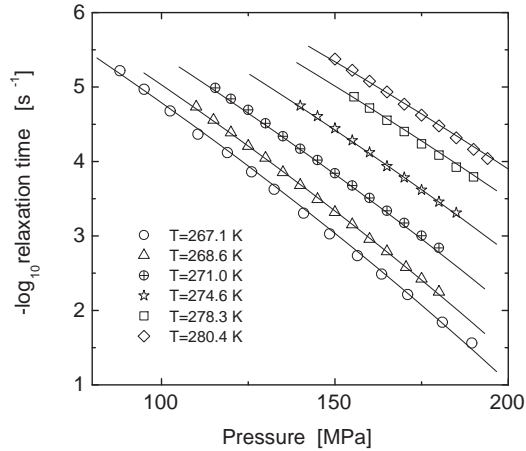}
\caption{\label{Fig_tau_PEAG} Structural relaxation time of salol
from photon-correlation measurements at different temperatures.
Data taken from Comez et al.\cite{ComezPRE02} ($T$=267.1 K
($\circ$), 268.6 K ($\vartriangle$), 271.0 K ($\oplus$), 274.6 K
($\star$), 278.3 K ($\square$), 280.4 K ($\lozenge$)). The
relaxation time is the average time $\langle\tau\rangle$. The
solid lines represent the simultaneous fit with the PEAG equation
--- Eq.~(\ref{AGphi}). As explained in the text, four parameters
are adjusted by the fitting procedure, in particular giving
$\left(\partial V/\partial T\right)_{0}^{non-struct}=(3.8\pm
0.7)\times 10^{-8}$ \mbox{m$^{3}$ mol$^{-1}$ K$^{-1}$} and
$\Phi=0.68\pm 0.08$.}
\end{figure}

\begin{figure}[t]
\includegraphics[width=8.2cm]{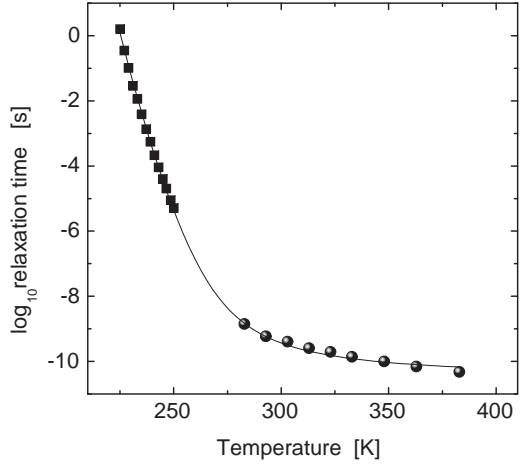}
\caption{\label{Fig_tau_CG0} Structural relaxation time of salol
from depolarized light scattering measurements at atmospheric
pressure. The relaxation time is the average time
$\langle\tau\rangle$. Squares represent depolarized
photon-correlation data from Ref.\onlinecite{Berger96}, circles
are depolarized Brillouin and Raman light scattering from
Ref.\onlinecite{Li92}. The solid line represents the fitting
curve using the CG equation
--- Eq.~(\ref{tauCG}). The four
parameters adjusted by the fitting procedure are $A_{CG}=(10.6\pm
0.1)$, $B_{CG}=(91\pm13)$K, $T_0=(265\pm3)$ K, and
$C_{CG}=(3.4\pm0.4)$ K.}
\end{figure}

\begin{figure}[t]
\includegraphics[width=8.2cm]{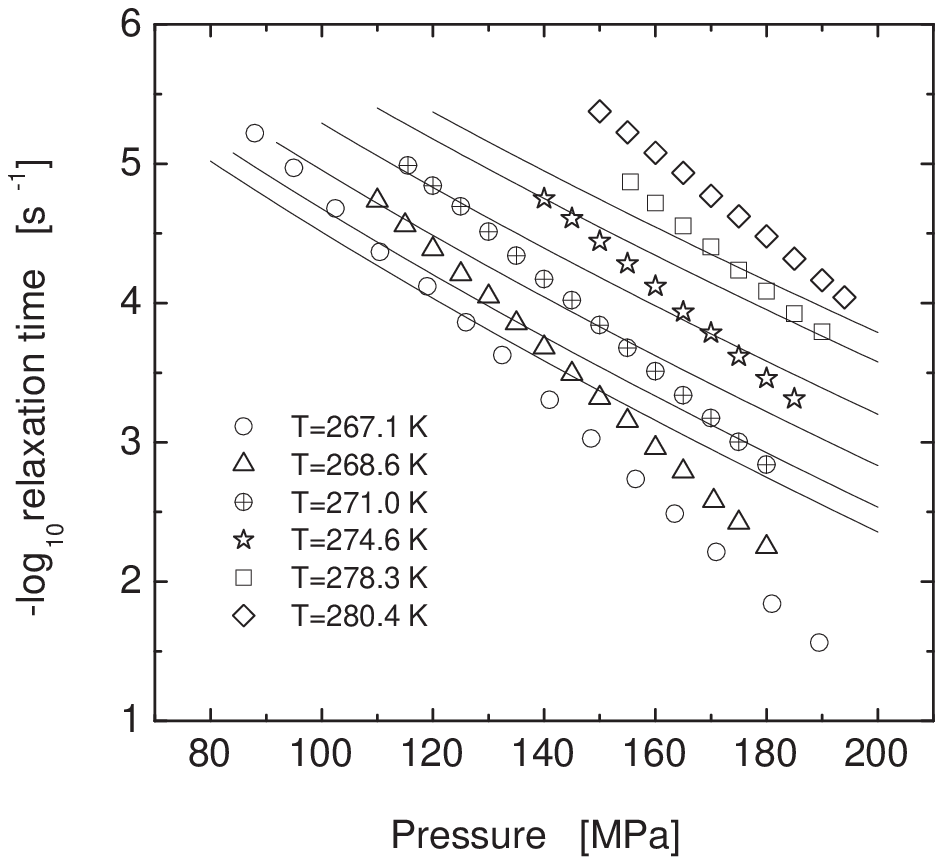}
\caption{\label{Fig_tau_CG} Structural relaxation time of salol
from photon-correlation measurements at different temperatures
\cite{ComezPRE02}. The relaxation time is the average time
$\langle\tau\rangle$. Temperatures are $T$=267.1 K ($\circ$),
268.6 K ($\vartriangle$), 271.0 K ($\oplus$), 274.6 K ($\star$),
278.3 K ($\square$), 280.4 K ($\lozenge$). The solid lines
represent the simultaneous fit with the pressure extended
Cohen-Grest equation --- Eq.~(\ref{tauCGex}). The parameters
$A_{CG}$, $B_{CG}$, $C_{CG}$, and $T_0$ have been taken fixed to
those obtained from the fit of the isobaric data at atmospheric
pressure.}
\end{figure}

Summarizing, in the fit of relaxation time data with
Eq.~(\ref{AGphi}) only four parameters, specifically $\tau_{0}$,
$C_{AG}$, $\Phi$, and $\left(\partial V/\partial T
\right)_{0}^{non-struct}$, remain to be adjusted. The fit is
carried out simultaneously in the $T$-$P$ domains, over the
pressure range $0.1-194$ MPa at six different temperatures
($T$=267.1, 268.6, 271.0, 274.6, 278.3, and 280.4 K). The best
fit curves (solid lines in Fig.~\ref{Fig_tau_PEAG}) correspond to
the values: $\log\tau_{0}[s]=-17.4\pm0.1$,
$C_{AG}=(1.9\pm0.3)\times 10^{5}$ \mbox{J mol$^{-1}$},
$\Phi=0.68\pm0.08$, $\left(\partial V/\partial T
\right)_{0}^{non-struct}=(3.8\pm 0.7)\times 10^{-8}$
\mbox{m$^{3}$ mol$^{-1}$ K$^{-1}$}.

It is important to remark that the value obtained for the
non-structural thermal expansion compares well with the value
calculated for the polycrystal of salol, while it is only in
feasible agreement with that estimated by some of us, $(\partial
V/\partial T)_{0}^{non-struct}=(1.09\pm 0.04)\times 10^{-8}$
\mbox{m$^{3}$ mol$^{-1}$ K$^{-1}$}, in a previous determination
using a preset $\Phi=1$ in Eq.~(\ref{AGphi}), i.e. obtained by
replacing the configurational entropy with the excess entropy
\cite{ComezPRE02}. Moreover, we note that the best fit yields a
value for $(\partial V/\partial T)_{0}^{non-struct}$ whose
uncertainty spans the variation with $T$ and $P$ of the crystal
thermal expansion. Thus, it emerges that the approximation
$(\partial V/\partial T)_{P}^{non-struct}\approx(\partial
V/\partial T)_{0}^{non-struct}$ is justified, and it is
unnecessary to consider a $T$ and $P$ dependence of the
non-structural expansion in Eq.~(\ref{Scisoth}).

\subsection{Check of the CG model}
\label{CheckCG}

Various models interpreting the dynamics of supercooled liquids
provide an equation to represent $\tau$ data as a function of
temperature. Among these, the most frequently used is the
Vogel-Fulcher-Tamman (VFT) equation \cite{Vogel}. However, its
adaptability to experimental data has been demonstrated only over
a limited range of temperatures. In fact, Stickel et al.
\cite{Stickel95,Stickel} have shown that two VFT equations are
needed to describe the relaxation data at ambient pressure for
temperatures ranging from just above the glass transition up to
very high temperatures, because of a change of dynamics occuring
in the vicinity of a crossover temperature $T_B\approx1.2T_g$. On
the other hand, the CG expression at ambient pressure, by virtue
of four characteristic parameters, one more than the VFT, succeeds
in describing structural relaxation times in a broad range of
temperatures. Positive tests have been reported on several glass
forming systems \cite{CohGre79,CumLi97,SchLun99}. Recently,
Paluch et al. \cite{Paluch03} have also shown that the
characteristic temperature $T_0$ of the CG model can be
identified with $T_B$ in a number of liquids, suggesting that the
change of dynamics may be related to an onset of percolation of
the free volume. However, estimates of the free volume available
per liquid-like molecule founded on such a description clash with
estimates extracted from dilatometric measurements \cite{Paluch03}.\\
An interesting and not frequently exploited testing-ground for
this model is the comparison with relaxation data obtained by
varying both temperature and pressure. To do this, in the case of
salol, we analyze the temperature dependent relaxation times at
ambient pressure, available in the literature
\cite{Li92,Berger96}, using Eq. (\ref{tauCG}), and compare the
results with those obtained from our data at variable pressure,
using Eq. (\ref{tauCGex}).

Depolarized light scattering measurements on salol performed at
ambient pressure by photon correlation spectroscopy
\cite{Berger96} and Brillouin and Raman spectroscopy \cite{Li92}
are reported in Fig. \ref{Fig_tau_CG0}, spanning a wide
time-temperature range. The fit parameters of Eq. (\ref{tauCG})
are: $A_{CG}=(10.6\pm 0.1)$, $B_{CG}=(91\pm13)$ K,
$C_{CG}=(3.4\pm0.4)$ K, and $T_0=(265\pm3)$ K, confirming that
$T_0$ matches the crossover temperature $T_B\simeq$265 K
\cite{Stickel,Paluch03}.

Then, we test the generalized CG equation on our $\tau(T,P)$
data. In the fit procedure, the parameters $A_{CG}$, $B_{CG}$,
$C_{CG}$, and $T_0$ are taken fixed to those obtained from the fit
of the data at ambient pressure, these four being the same
parameters which also appear in Eq. (\ref{tauCG}), and $\xi_0$ is
the only free parameter. The inability of the CG equation to
represent the variation of $\tau$ with pressure is apparent in
Fig. 6, where the solid lines are generated by Eq.
(\ref{tauCGex}). On the other hand, treating all the parameters
as free the fitting algorithm does not converge. A similar result
has also been obtained for an epoxy system \cite{CorezziCPL}. The
inapplicability of the generalized CG equation prompts disfavor
towards the robustness of the CG theory. Nevertheless, the free
volume approach remains physically attractive, and we cannot
exclude that the inadequacy of Eq. (\ref{tauCGex}) to describe
the $\tau(T,P)$ data might be ascribed to the number of
simplifications used to derive the equation, which are possibly
no longer valid at high pressures.

\section{Conclusions}
\label{conclusions}

In conclusion, we have studied the slow dynamics of salol under
variable temperature and pressure using PCS in combination with
PVT measurements. Comparing the behavior of the structural
relaxation time with equations derived within the AG entropy
theory and the CG free volume theory, we find that pressure
dependent data are crucial to assess the validity of model
equations of the glass transition. In particular, we confirm
previous work \cite{CorezziCPL} showing that the pressure
dependent expression of $\tau$ predicted by the CG model cannot
reproduce the experimental data, despite the presence of five
adjustable parameters and an ability to parametrize $\tau(T)$
data over a broad temperature range at ambient pressure. Instead,
experimental $\tau(T,P)$ data conform to the entropy-based PEAG
equation. Interestingly, since the parameters which control the
pressure dependence of $\tau$ have separately been determined via
PVT measurements, this equation requires only four adjustable
parameters in the $T$ and $P$ intervals investigated in the
present work. Remarkably, the deduced parameters yield physical
results. Especially, the fraction of excess entropy which arises
from structural configurations is realistically estimated ($\sim$
70\% at ambient pressure).\\
In an effort to determine the role played by volume and thermal
effects in driving molecular dynamics, Casalini et al.
\cite{CasaliniJPC03} have recognized that neither temperature nor
volume is the dominant variable governing the structural
relaxation of salol near $T_{g}$, consistently with results for a
number of other glass formers \cite{PaluchJCP03b,PaluchPRB2002}.
Conceptually, this result accords with our findings that the
dominant thermodynamic variable is configurational entropy, a
quantity which embodies both temperature and volume effects:
different relative contributions to $\tau$ of thermal energy and
volume reflect a different sensitivity of the number of
configurations to change following temperature and volume changes.

We believe that the positive test of the PEAG model presented
here should stimulate further work on other glass formers and by
different techniques.

\medskip

We thank Prof. E.W. Fischer and Prof. C.A. Angell for valuable
comments and suggestions .


\begin{references}
\bibitem{expansion} Here we use the notation $\left(\partial V/\partial T \right)_{P}$
for thermal expansion, and $1/V\!\left(\partial V/\partial T
\right)_{P}$ for thermal expansion coefficient.
\bibitem{zeropressure} For the present purposes, the atmospheric
pressure ($P$=0.1 MPa) can be considered as $P\approx0$.
\end{references}
\end{document}